\documentclass[twocolumn,showpacs,prl,superscriptaddress,floatfix]{revtex4}
\usepackage[toc]{appendix}
\usepackage{color} 
\usepackage{tabularx} 	
\usepackage{epsfig}
\usepackage{amsmath} 
\usepackage{amssymb} 
\usepackage{graphicx}
\usepackage{times}
\usepackage{comment}

\newcommand{\beq}{\begin{equation}}
\newcommand{\eeq}{\end{equation}}
\newcommand{\la}{\langle}
\newcommand{\ra}{\rangle}
\newcommand{\e}{\varepsilon}

\newcommand{\uss}{\mathfrak{U}} 

\begin{document}

\title{Avalanches in the Relaxation Dynamics of Electron Glasses}

\author{Martin Goethe}
\affiliation{Departament de Qu\'imica Inorg\`anica i Org\`anica,
Universitat de Barcelona,  Mart\'i i Franqu\`es 1, E--08028 Barcelona, Spain.}
\affiliation{Departament de F\'\i sica de la Mat\`eria Condensada,
  Universitat de Barcelona, Mart\'i i Franqu\`es 1, E--08028 Barcelona, Spain.}
\author{Matteo Palassini}\email{palassini@ub.edu}
\affiliation{Departament de F\'\i sica de la Mat\`eria Condensada,
  Universitat de Barcelona, 
Mart\'i i Franqu\`es 1, E--08028 Barcelona, Spain.}
\affiliation{Universitat de Barcelona Institute of Complex Systems (UBICS), 
Mart\'i i Franqu\`es 1, E--08028, Barcelona, Spain}
\affiliation{CIBER-BBN Center for Bioengineering, Biomaterials and Nanomedicine, Instituto de Salud Carlos III, Madrid, Spain}
\date{\today}

\begin{abstract}
We study the zero-temperature relaxation dynamics of
an electron glass model with single-electron hops. We find numerically that
in the charge rearrangements (avalanches) triggered by displacing an electron, the number 
of electron hops has a scale-free, power-law distribution up to a cutoff 
diverging with the system size $N$, independently of the disorder strength
and provided hops of arbitrary length are allowed. 
In avalanches triggered by the injection of an extra electron,
the distribution does not have a power-law limit, but its 
mean diverges non-trivially with $N$. 
In both cases, the avalanche statistics is
well reproduced by a branching process model, that
assumes independent hops. Qualitative
differences with avalanches in infinite-range spin glasses 
and related systems are discussed.

\end{abstract}

\pacs{72.80Ng, 64.60.av, 75.10.Nr }

\maketitle

The long-range Coulomb interaction plays a prominent role 
in disordered systems of localized electrons known as electron glasses
\cite{ESbook,OrtunoBook,PalassiniCS}.
 Its best known manifestation is the Coulomb (pseudo)gap, namely the 
vanishing of the single-particle density of states (DOS) 
as a power law  $g(\e)\simeq |\e-\mu|^\delta$ near the chemical 
potential $\mu$ \cite{ES75,Pollak}.
This led to the prediction of a modified hopping conductivity  
\cite{ES75,ESbook}, confirmed experimentally in many 
disordered insulators \cite{OrtunoBook}.

The similarity between models of electron glasses and 
disordered spin systems inspired the conjecture of 
a transition to a spin-glass-like equilibrium  phase
\cite{Davies},
analogous to the Almeida-Thouless transition \cite{AT} in the infinite-range
Sherrington-Kirkpatrick (SK) spin glass model \cite{SK}.
No sign of this transition, later also predicted by replica mean-field 
theory \cite{glass, pastorPRL}, was found in Monte Carlo studies down 
to very low temperatures \cite{gp,Surer}. Its absence would
not be incompatible with the ample experimental 
evidence of glassy nonequilibrium dynamics
in disordered insulators, the
origin of which is still not well understood
\cite{BenChorin, Grenet,Frydman,Ovadyahu_review, OrtunoBook}.

The SK model does share with electron
glasses two features: {\em i)} a rugged free energy landscape with 
a multitude of metastable states, induced by frustration;
{\em ii)} a pseudogap in the distribution of local fields, 
analogous to the Coulomb gap, both in 
metastable states \cite{Palmer, Pazmandi} and in the ground state
\cite{TAP}. As a result, its zero-temperature relaxation dynamics
 exhibits {\em crackling}:
a quasistatic change of the external field induces large rearrangements
(``avalanches'') with a scale-free size distribution
 $p(S)\sim S^{-\tau}$ up to a cutoff that diverges with the system
size $N$, with $\tau\simeq 1$ 
\cite{Pazmandi, MW, pastor} and without any parameter fine tuning
(here and below, $S$ is the number of elementary relaxations, in this case 
single spin flips, during an avalanche).
Such ``self-organized criticality'' \cite{Pazmandi} contrasts with
short-range Ising spin glasses, which lack both a pseudogap \cite{boettcher}
and a scale-free $p(S)$ \cite{goncalves,andresen2},
and the (short-range) random-field Ising model, in which a power-law $p(S)$  
is observed only at a critical point \cite{RFIM}.

Scale-free avalanches occur in vortices 
in type-II superconductors 
\cite{Field}, martensitic transitions \cite{Vives}, 
 earthquakes, and many other systems, and their
universality is currently debated \cite{Dahmen}.
In systems driven quasistatically by an external
field, it was predicted that if the elementary 
excitations in the relaxation dynamics have a pseudogap, 
the mean avalanche size $\langle S \rangle$ diverges as a power
of $N$ dictated by the pseudogap exponent \cite{MW}.
This scenario was applied to the SK model (see also Ref.~\cite{Pazmandi}), 
dense granular and suspension flows near jamming, and the plasticity
of amorphous solids \cite{MW}.

In this paper, motivated by the analogy with the SK model
and by experimental hints of avalanche-like behaviour in
disordered insulators \cite{Monroe, Ovadyahu2014}, we investigate 
avalanches in the relaxation dynamics of
electron glasses with single electron hops.
We show numerically that, if hops of arbitrary length are allowed,
in {\em displacement avalanches} (triggered by moving an electron)
$p(S)$, where $S$ is the number of hops, tends to a power-law with
an exponent consistent with the mean-field value $\tau=3/2$.
In {\em injection avalanches}  (triggered by adding an electron to
the softest empty site,
the equivalent to quasistatically changing the field in an Ising spin system)
$\langle S \rangle$ diverges with $N$, but
$p(S)$ does not tend to a power-law in the thermodynamic limit.
We propose a branching process model that reproduces well the 
observed statistics in both cases, and predicts 
that the scaling  of $\langle S \rangle$ with $N$
is not governed by the exponent $\delta$, which however enters 
in the scaling with the disorder strength. 
These findings indicate that the avalanche process is 
qualitatively different from that of the SK model and of 
a related artificial electron glass dynamics 
with non-conserved number of electrons \cite{MW},
reflecting the different nature of the elementary excitations, which
in the electron glass are electron-hole pairs lacking a pseudogap. 
A brief account of some of our results 
has appeared in Ref.\cite{acre}.

We study the Efros model  \cite{efros76} with classical Hamiltonian
\beq
\mathcal H =  \frac{e^2}{2}\sum_{i\neq j}(n_i - K) \frac{1}{r_{ij}} (n_j - K)
+ \sum_i n_i \varphi_i \, ,
\label{ES}
\eeq
where $n_i \in \{0,1\}$ are the occupation numbers for the $N=L^d$ sites  
of a cubic (square) lattice of linear size $L$ for $d=3$ ($d=2$),
$r_{ij}$ is the distance between $i$ and $j$, $\varphi_i$ are 
independent, Gaussian-distributed variables with zero mean and standard
deviation $W$,
and $e$ is the electron charge 
divided by the square root of the lattice dielectric constant. 
Neutrality is ensured by imposing $\sum_{i=1}^N n_i = K N$,
and periodic boundary conditions are implemented with an Ewald summation. 
Numerical values of 
distances (energies) are given in units of the lattice spacing 
$a$ ($e^2/a$).
The dynamics consists of energy-lowering single-electron
hops $i\to j$ with a transition rate $\Gamma_{ij} \propto \exp(-2 r_{ij}/\xi)$,
where $\xi$ is the localization length. We consider two opposite limits: 
$\xi \to 0$, whereby hops can relax only the shortest 
unstable electron-hole pairs available at any given time, 
and  $\xi \to \infty$, whereby 
 unstable pairs of all lengths are equally likely to relax 
(this notation does not imply delocalization).

The system is prepared in a 
metastable state (a configuration stable against all single-electron
hops) by quenching it instantaneously to zero temperature from
a random configuration, and evolving it with the $\xi=\infty$ 
dynamics (which allows to find metastable states efficiently
\cite{Glatz_2008})  until no unstable pairs are left. 
Next, we perturb the system by either displacing or injecting an electron.
This usually destabilizes some pairs, which upon relaxing 
(under the $\xi=0$ or the $\xi=\infty$ dynamics) can destabilize
other pairs, creating an avalanche 
that stops   when a new metastable state is reached after $S$ hops.
The procedure is repeated for many ($\simeq 10^3$ to $10^5$) 
samples with different $\{\varphi_i\}$.

For displacement avalanches in 3D
under the $\xi=\infty$ dynamics, the avalanche size distribution 
is well fitted by a power-law with a cutoff 
proportional to the linear size $L$,
\beq
p(S) = \frac{A_L}{S^\tau} e^{-S/S_c}, \quad S_c = a L
\label{pS}
\eeq
with $\tau = 1.4(1)$, as shown in Fig.1 for $W=2$, $K=1/2$,
which extrapolates to a scale-free form in the thermodynamic limit.
\begin{figure}
\begin{center}
  \includegraphics[height=\linewidth,width=\linewidth,angle=270]{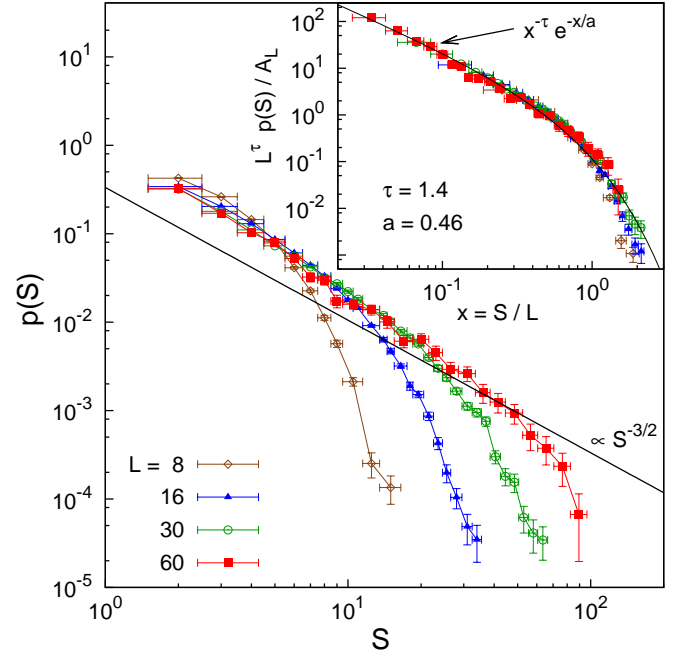}
\caption{(Color online) Size distribution of the avalanches triggered by
an electron displacement and evolved with the $\xi=\infty$ dynamics in 3D, 
for $W=2$, $K=1/2$. $p(S)$ is obtained from ($90k,48k,29k,2.1k$) samples 
for $L=(8,16,30,60)$. The initial displacement is the hop of
minimal energy that destabilizes at least another pair. 
Inset: scaling plot of the same data, according to Eq.(\ref{pS}), 
with $A_L$ fixed by normalization.
}
\label{Fig1}
\end{center}
\end{figure}
\begin{figure}
\includegraphics[height=\linewidth,angle=270]{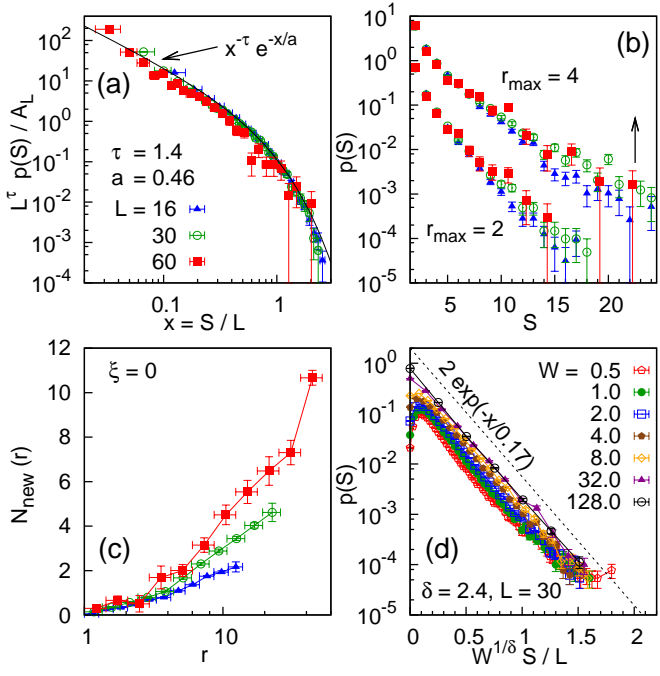}
\caption{(Color online) {\bf (a)} Same as the inset of Fig.1 but for 
the $\xi=0$ dynamics. The initial displacement is the hop of length $r_{ij}=1$
with the smallest energy, which is never
allowed to relax back, so the final state is not necessarily
metastable. This slight difference from the protocol of Fig.1 is unimportant.
{\bf (b)} Same as Fig.1 but for the $\xi=0$ dynamics
with bounded hopping length $r\leq r_{max}$. 
{\bf (c)} Average number of pairs destabilized
after a hop of length $r$ for different system sizes. Protocol and
symbols are as in panel (a). 
{\bf (d)} Rescaled size distribution for injection avalanches 
in 3D evolved with $\xi=\infty$, for 
 different disorder strength $W$, $L=30$, and $K$ in the range
$[\frac{1}{2}-\frac{1}{\sqrt{N}},\frac{1}{2}+ \frac{1}{\sqrt{N}}]$.}
\label{Fig2}
\end{figure}
Before addressing the origin of the scale-free behavior, it is useful
to analyze the elementary excitations in the avalanches.
The energy change due to a hop $i\to j$ is
$\omega_{ij} = \e_j - \e_i - e^2/r_{ij}$
where
$\e_i = e^2 \sum_{j \neq i} (n_j - K) / r_{ij} + \varphi_i$
is the energy required to add an electron at site $i$, leaving
the rest of the system unchanged. 
The pairs destabilized in the avalanche
are ``soft'', with length 
$r_{ij}$ comparable to $e^2/|\e_j-\e_i|$.
We find that most are already soft in the initial state
(for example, the fraction of hops
for which $\omega_{ij} > 0.1$ in the initial state
is $0.27, 0.17, 0.12$ for $L=16,30,60$). Thus, the avalanche involves
only a small fraction of the sites, as most pairs are ``hard''.
Efros and Shklovskii argued that 
$g(\e)=\langle\delta(\e_i-\e)\rangle$ vanishes 
as $|\e-\mu|^\delta$ for $\e\to \mu$, with 
$\delta \geq d-1$ \cite{ES75}, forming a Coulomb gap
of width $\Delta \sim W^{-1/\delta}$. This was confirmed in 
numerical computations that give
$\delta \simeq 2.4\div 2.7$ in three dimensions \cite{acre,mobius},
possibly with a crossover to exponential behavior at 
very low energies \cite{baranovski80, skinner},
and $\delta \simeq 1.0 \div 1.2$ in two dimensions \cite{mobius,li,unpublished}.
The vast majority of soft pairs are thus {\em dipoles}
formed by two hard sites with $|\e_{i,j}-\mu| >\Delta$,
with typical length
$r_{ij}\simeq r_0 = e^2/\Delta$. 
Long soft pairs with $r_{ij}\gg r_0$ involve sites inside the Coulomb gap
($|\e_{i,j}-\mu|\ll \Delta$)
and thus are much rarer. 
Neglecting  electron-hole correlations, 
the pair DOS was estimated to 
behave as $(\omega + 1/r)^{2 \delta +1}$ for $r\gg r_0$
and $\omega\ll \Delta$
\cite{baranovski80}. We find 
that the length distribution of the hops performed during
an avalanche decays as $r^{-5.3}$ in 3D, as shown in 
Fig.S2 of the Supplemental Material \cite{Supp},
in fairly good agreement with this estimate. 
The preponderance of short hops even in the $\xi=\infty$ dynamics
explains why $p(S)$ maintains an almost identical shape 
if we let the avalanches evolve with the $\xi=0$ dynamics 
instead, as shown in Fig.2a.

Let us now turn to injection avalanches, created by adding an extra
electron at the empty site with the smallest 
$\e_i>0$, and simultaneously incrementing the background charge $K$ by $1/N$, to
preserve neutrality and avoid the
infinite energy required to charge the periodic lattice.
Unlike in displacement avalanches, $p(S)$ now has a maximum
(Fig.3), which can be understood at mean-field level if we assume
that the dipoles destabilized by the 
initial injection are uncorrelated and thus their number $M$ is 
Poisson distributed. Then $p(S=0)=e^{-\langle M\rangle}$, 
and $\langle M \rangle$ can be estimated 
by counting the pairs with energy smaller than the charge-dipole interaction,
\beq
\langle M \rangle \simeq \int_a^L dr \,r^{d-1} \int_0^{\frac{e^2 r_0}{r^2}} d\omega \Phi(\omega)
 \simeq r_0 e^2 g_0 L^{d-2} \, ,
\label{muL}
\eeq
where the total pair DOS, $\Phi(\omega)=\langle \delta(\omega_{ij}-\omega)\rangle$, 
for $\omega\ll \Delta$ is of order of the bare single-particle DOS, $g_0=e^2/(a^d W)$
(apart from logarithmic corrections in 3D) \cite{baranovski80}.
Thus $\langle M\rangle \propto L W^{(1-\delta)/\delta}$ for $d=3$ and
$p(S=0)$ decreases exponentially in $L$, consistent with the data in Fig.3.         
As shown in the inset of Fig.3, the exponential cutoff in the tail of $p(S)$ 
increases linearly with $L$ as in Eq.(\ref{pS}), with $\tau = 1.5(1)$. 

Since dipoles are separated from each other by a distance much larger than $r_0$,
at zero temperature $r_0$ and $L$ are the only length scales 
in avalanches dominated by dipoles. Hence the cutoff, in both types of avalanches, 
should depend on $L$ via the ratio $L/r_0$, a scale invariance stemming
entirely from the Coulomb gap. We confirm this prediction by plotting $p(S)$ against 
$S / (L W^{-1/\delta})$ in Fig.2d: 
the tails for all values of $W$ 
(away from the charge-ordered phase \cite{gp}) 
collapse onto the same slope for $\delta = 2.4$, 
in excellent agreement with an independent estimate of $\delta$
from the shape of the Coulomb gap \cite{acre}. 
In injection avalanches, $\langle M\rangle$ sets an additional 
characteristic scale in $p(S)$ that also diverges linearly in $L$, but 
does not scale with $r_0$ since it depends on the charge-dipole 
rather than the dipole-dipole interaction. We argue later that, as a consequence,
$p(S)$ does not develop a power-law tail for $L\to \infty$.

\begin{figure}
\includegraphics[height=\linewidth,angle=270]{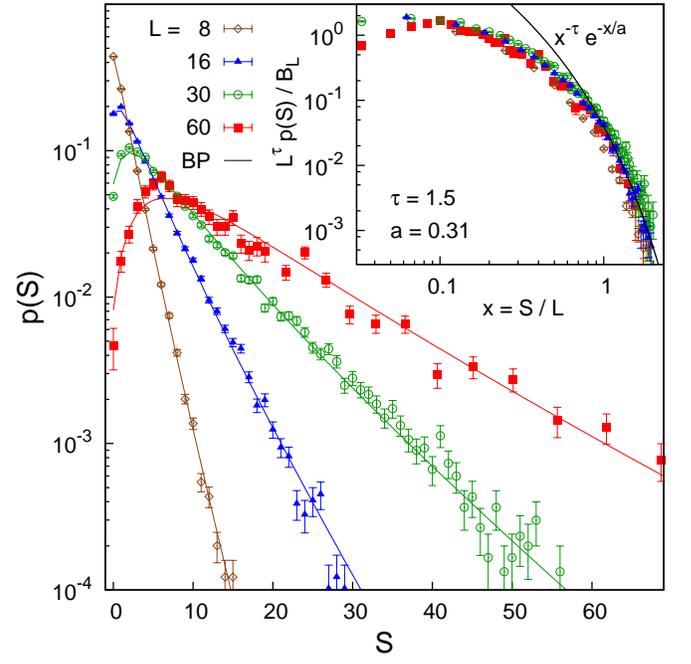}
\caption{(Color online) Same as Fig.1 but for avalanches triggered 
by injection. The same samples as in Fig.1 were used. 
The lines in the main figure represent the best fit,  for each value of $L$,
to branching process analytical result,
Eq.(S11) of the Supplementary Material \cite{Supp}. In the inset, $B_L$ is
fixed by normalization, see Eq.(S12) in Ref.\cite{Supp}.} 
\label{FigCTA}
\end{figure}

In 2D, we find that in both displacement and injection avalanches 
the tail of $p(S)$ is well fitted by Eq.(\ref{pS}),
but with a cutoff diverging logarithmically in $L$ (see 
Fig.S5 and S6 in \cite{Supp}). Moreover, for injection avalanches
$p(S=0)$ decreases with $L$ much more slowly than in 3D,
in agreement with the estimate
 $\langle M\rangle \propto W^{(1-\delta)/\delta} \log (L/a)$ that follows from Eq.(\ref{muL}).

What is the origin of large avalanches? An argument of M\"uller and Wyart \cite{MW}, 
adapted to the electron glass (see \cite{Supp}, Section B),
predicts that if single electrons can be exchanged with a reservoir 
from anywhere in the bulk, the mean net number of electrons exchanged
during an injection avalanche diverges as $\Delta N_e \sim N^{\delta/(1+\delta)}/W$.
Hence $\langle S \rangle \geq L^{d-1}$, as follows from
$\delta \geq d-1$ and $S\geq \Delta N_e$. 
As observed in Ref.\cite{MW}, 
this dynamics is not realistic (in a real system exchanges 
can only occur at the sample leads), but
bulk single-site exchanges are essential for their argument,
which holds
also if hops between sites are not allowed, since the single-site 
exchanges suffice to create a Coulomb gap with $\delta\geq d-1$ \cite{foot}.

The argument of Ref.\cite{MW} does not apply to injection
avalanches under the
particle-conserving dynamics considered here,
nor to displacement avalanches with or without single-site exchanges.
A key observation to understand particle-conserving large
avalanches is that long hops, even if rare,  are 
necessary to sustain them. To see this, we modified
the $\xi=0$ dynamics by imposing a maximum
allowed hopping length $r_{max}$ independent of $L$,
and found that the exponential cutoff $S_c$
now does not change with $L$, as shown in Fig.2b for displacement avalanches
(similar data were obtained in Ref.\cite{andresen}). 
This agrees with the converse of the argument of Ref.\cite{MW}, which
shows that in the absence of a pseudogap the mean avalanche size remains
finite. In fact, the only active excitations are now
dipoles with $e^2 r_0^2/r^3$ interaction that do not have
a power-law pseudogap.
If however we set $r_{max}=L/10$, 
the cutoff in $p(S)$ becomes again proportional to $L$ (see Fig.S1 
in \cite{Supp}).
The importance of long hops stems from their larger destabilizing effect, 
shown in Fig.2c, as they act as single-site excitations near the two sites 
involved, and are more likely to affect regions still 
untouched by the avalanche.

Motivated  by the closeness of $\tau$ to the mean-field value $3/2$ \cite{RFIM},
we model the avalanches as a branching process,
assuming that every hop creates a random number $X$ of subsequent hops 
with  mean $\langle X \rangle=\lambda$.
Then, displacement avalanches are described by the well-known
Galton-Watson (GW) process \cite{Harris}:
if $\lambda>1$,  ``explosive'' avalanches with $S=O(N)$ occur with
finite probability; if $\lambda < 1$, one has $\langle S\rangle=(1-\lambda)^{-1}$
and $p(S)$ decays as in Eq.(\ref{pS}) with $\tau=3/2$, with a cutoff diverging as
$S_c\sim \langle X^2\rangle/(1-\lambda)^2$ as $\lambda\to 1^-$
 \cite{Harris, Fisher}. Assuming that $X$ is Poisson distributed,
$S_c$ is a function of $\lambda$ only, and from the fit $S_c=0.46 L$ 
in Fig.2  we obtain an effective $L$-dependent $\lambda$ that 
tends to one for $L\to \infty$ and agrees well,
for each $L$, with the mean $\langle X\rangle$ measured 
by reconstructing the tree of events in each avalanche \cite{Supp}.

We model injection avalanche as the compound of $M$ independent 
``sub-avalanches'', each triggered by the relaxation of
one the dipoles destabilized by the injection. Hence
$S=\sum_{m=1}^M S_m$, where $S_m$ is the size of a sub-avalanche.
Assuming that the tail of the distribution of $S_m$ has the form in 
Eq.(\ref{pS}), we have
$\langle S \rangle \sim \langle M \rangle S_c^{2-\tau}$, and thus Eqs.(\ref{pS}) 
and (\ref{muL}) imply $\langle S\rangle \sim L^{3-\tau}/W^{(\delta+1-\tau)/\delta}$ in 3D,
in agreement with our numerical data (not shown). This differs from the scaling
of $\langle S\rangle$ with $L$ and $W$ in the non particle-conserving dynamics,
mentioned earlier. Assuming that the sub-avalanches are GW processes and 
$X$, $M$ are Poisson distributed (as supported by our data \cite{Supp}), using
a generating function method we obtain an analytical expression for $p(S)$
(Eq.(S11) in Ref.\cite{Supp}) parametrized by $\langle M \rangle$ and $\lambda$, which for
$S\gg \langle M\rangle$ has the asymptotic form 
$p(S)\sim \langle M\rangle S^{-3/2} \exp(-S/S_c(\lambda))$ with  the same 
dependence $S_c(\lambda)$ as in diplacement avalanches.
The full analytical expression fits very well our data, as shown in Fig.3.
The fits are consistent with a linear scaling 
of $\langle M \rangle$ and $S_c(\lambda)$ in $L$, and agree fairly well
for each $L$ with the
values of $\langle M\rangle$ and $\lambda$ that obtained
from the avalanche tree reconstruction \cite{Supp}.
We stress that the analytical expression for $p(S)$  does not have a power-law tail in
the thermodynamic limit since the position of its maximum $S^*$
increases with $L$ faster than $S_c$ does ($S^*/S_c \sim L^{1/2}$).
This implies that the mean $\langle S\rangle$ is not
dominated by rare events, unlike in displacement avalanches, 
and it explains why the scaled data 
in the inset of Fig.3 do not show a region with constant slope. 
We thus have crackling (diverging mean avalanche size) without
a power-law distribution.
Further details and tests of the branching process 
model are discussed in Ref.\cite{Supp}. Despite neglecting the 
correlations between hops, it reproduces remarkably
well the avalanche statistics,
suggesting that the system self-organizes to reach the critical 
point $\lambda = 1$. Finding a dynamical explanation of how it does so is a challenging
task. 

To conclude, our results show that,
due to a combination of long-range interaction and long hops,
electron glasses displays 
a self-organized crackling that is qualitatively different 
from that of the SK model, and is well captured by a mean-field description.
To assess the experimental implications of these results,
one need to address the time scales of the avalanches
as well as the role of temperature
and multi-electron transitions. The typical length $r_{typ}$ of thermally activated
hops should act as a soft cutoff on the hopping length, and thus on
avalanche sizes. Based on the percolation approach to 
variable-range hopping for the model considered here \cite{ESbook},
we estimate \cite{example} that large avalanches could be observable 
at reasonable time scales at low temperatures. 
It is likely that multi-electron transitions will contribute significantly 
to the relaxation before reaching these time scales, as observed in 2D
simulations \cite{bergli}, speeding up the approach to
 equilibrium, which should shrink further the avalanches.
To quantify the impact of such transitions remains a much debated problem
\cite{Somoza2006, Pollak2018}.
It would be interesting to search for large rearrangements in disordered
insulators in charging experiments \cite{Orihuela}, or via their effect on the
percolating paths in conduction experiments \cite{Ovadyahu2014}.

We thank Markus M\"uller for discussions. 
This work is supported by AGAUR (2014SGR1379)
and MINECO (FIS2015-71582-C2-2-P). 
We thankfully acknowledge the computer resources at RES-BSC.

\begin{appendices}

\renewcommand{\theequation}{S\arabic{equation}}
\setcounter{equation}{0}
\setcounter{figure}{0}
\renewcommand{\thefigure}{S\arabic{figure}}



\date{\today}
\thispagestyle{empty}
\vspace{2cm}

\def\thesection{\Alph{section}} 
\def\thesubsection{\arabic{section}.\arabic{subsection}} 
\def\thesubsubsection{\arabic{section}.\arabic{subsection}.\arabic{subsubsection}} 
\def\thefigure{S\arabic{figure}} 
\newpage


\begin{center}{{\bf Supplemental material: Avalanches in the Relaxation Dynamics of Electron Glasses}}
\end{center}
\section{A. Additional details for simulations in three dimensions}

\begin{figure}[h!]
\includegraphics[width=7cm,angle=-90]{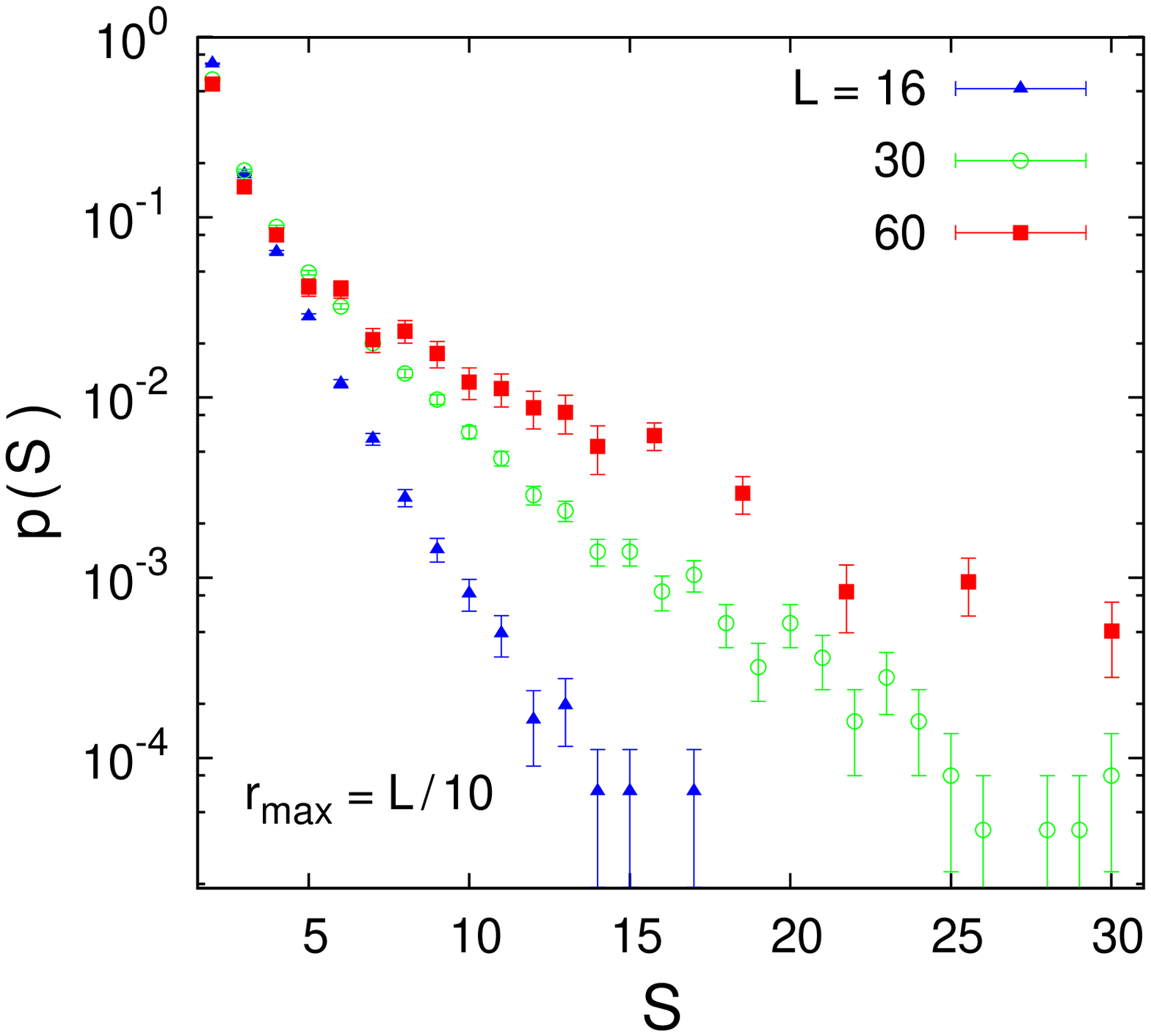}
\caption{Size distribution of avalanches triggered by
an electron displacement and evolved with the $\xi=0$ dynamics in 3D with 
hopping length restricted to $r\leq r_{max}=L/10$, for $W=2$, $K=1/2$. 
}
\label{fig:restricted}
\end{figure}

Fig.\ref{fig:restricted} shows the size distribution of displacement avalanches
under the $\xi=0$ dynamics, restricting the hopping length to
$r \leq r_{max} = L/10$. In this case, the exponential cutoff increases
with the linear system size $L$, unlike in the case of constant $r_{max}$ shown in Fig.2b of the main
text, in which the cutoff is independent of $L$.

\begin{figure}[h!]
\includegraphics[width=7cm,angle=-90]{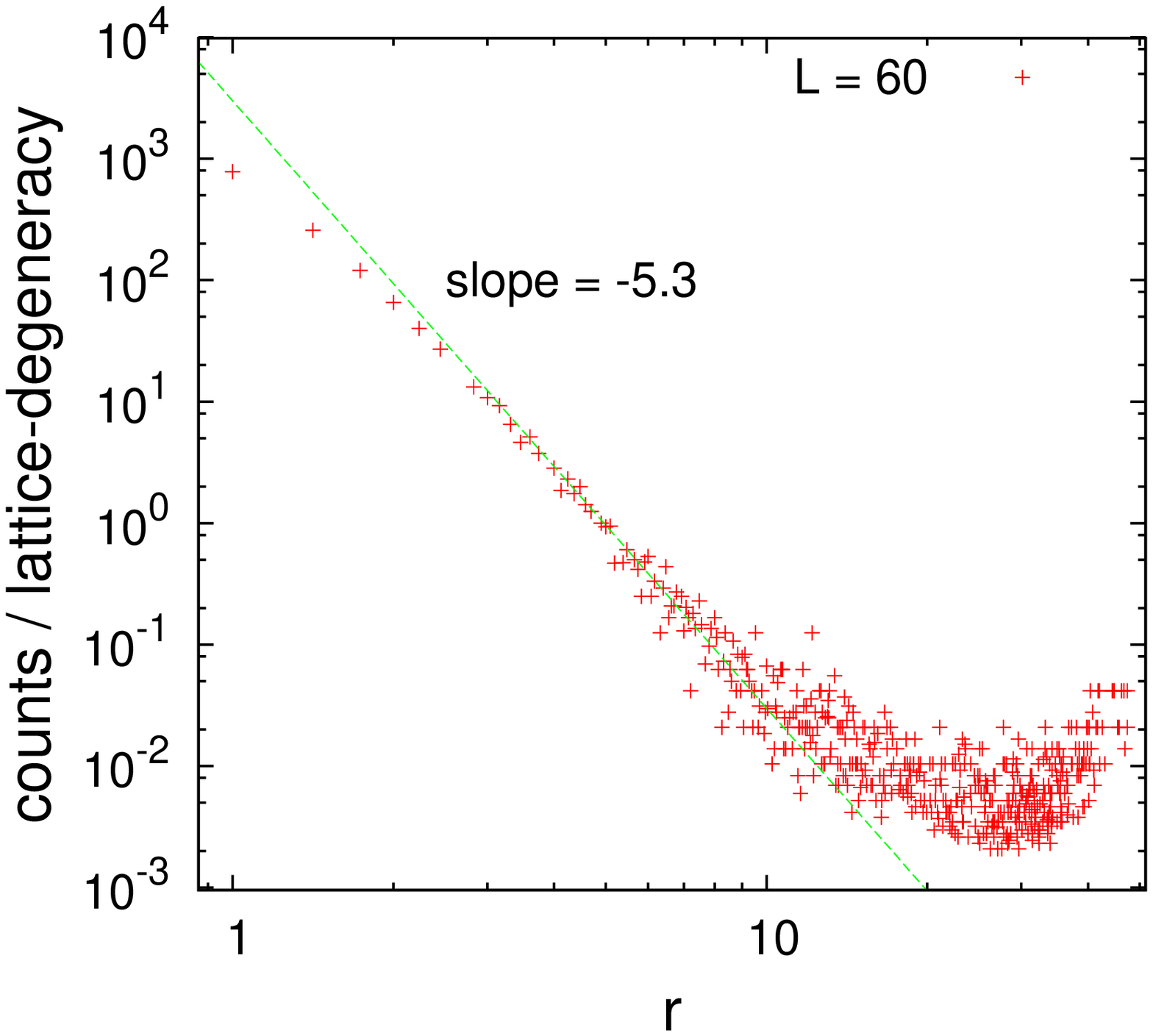}
\caption{Unnormalized distribution of the length of the hops taking place during the course 
of avalanches triggered by an electron displacement. The quantity displayed is the number of
hops of length $r$, divided the lattice degeneracy
(i.e. the number of lattice sites at distance $r$ from 
a given site, in a periodic cubic lattice of size $L=60$.)}
\label{fig:length_dis}
\end{figure}

The length distribution of the hops taking place during a displacement avalanche
is shown in Fig.\ref{fig:length_dis}.
 The data up to
$r\simeq 10$ lattice spacings are consistent
with the power-law decay of the distribution as 
$\approx 1/r^{2 \delta + 1}$, in agreement with
a mean-field argument based on the Coulomb gap \cite{baranovski80} for the
DOS of electron-hole pairs. The deviation at large $r$ is a finite-size effect.

\section{B. Mean avalanche size for non-particle-conserving dynamics}

We reproduce here the argument of
M\"uller and Wyart \cite{MW}, adapted to 
injection avalanches in the electron glass. The
argument assumes a non particle-conserving dynamics in which
single
electrons can be added or removed at any site. 
In addition, particle-conserving moves, such as single-electron
hops, can be allowed but are not necessary for the following argument.

Consider avalanches 
created by raising the chemical potential $\mu$ quasistatically until one
site becomes unstable, which is equivalent to adding an electron
at the softest site in our protocol.
We assume that the dynamics is such that after each avalanche a Coulomb gap is restored, 
so that the single-particle DOS behaves as $g(\epsilon)=c_d |\epsilon-\mu|^\delta$ 
for small $|\epsilon-\mu|$. As mentioned in the main text, single-site exchanges are
sufficient to restore a Coulomb gap with $\delta\geq d-1$.
Then, at each electron injection the chemical potential 
$\mu$ is shifted by an amount $\Delta \mu$ such that $N a^d \int_\mu^{\mu+ \Delta \mu} g(\epsilon) d\epsilon$ is
of order one, i.e.
$\Delta \mu \sim (c_d a^d N)^{-1/(\delta +1)}$.
Provided neutrality is maintained by changing the compensating charge $K$ at
each particle exchange, 
 the inverse compressibility $\chi(\mu)^{-1}=d\mu/d\langle K\rangle$ remains finite.
Consider now successive avalanches created by sweeping quasistatically 
$\mu$ in a range $[-\mu_{max},\mu_{max}]$, 
with $\mu_{max}$ of order of the disorder strength $W$ so that $K(\mu_{max}) \simeq 1$
(and $K(-\mu_{max})\simeq 0$).
The average net number of electrons 
entering the system at each avalanche is 
$ \Delta N_e = N \langle \chi(\mu) \rangle \Delta \mu
$,
and since $\langle \chi(\mu) \rangle \simeq 1/W$, we obtain
\begin{equation}
\Delta N_e \simeq N^{\delta/(1+\delta)}/[W (c_d a^d)^{1/(\delta+1)}]
\label{deltaNe}
\end{equation}

Two observations not included in the argument of Ref.\onlinecite{MW} are in order. First, if the charge is not compensated by changing $K$, then the system is incompressible ($\chi(\mu)^{-1}$ diverges with $L$), so the above argument does not hold. 
Second, the finite system size modifies the shape of the Coulomb gap for $|\epsilon-\mu|$ below an energy of order $e^2/L$, and the above argument needs to be modified accordingly.
Because of this finite-size effect, $\Delta \mu \sim e^2/L$, with
a proportionality constant that depends on the boundary conditions. Thus Eq.(\ref{deltaNe}) gets modified as  
$\Delta N_e \simeq e^2 L^{d-1}/(W a^d)$. This is the same form 
taken by Eq.(\ref{deltaNe}) if the Efros-Shklovskii bound is saturated, i.e. $\delta = d-1$ (in which case $c_d$ is proportional to $e^{-2d}$),
which was found numerically to be the case when the only moves allowed are the particle exchanges with the reservoir \cite{unpublished}.

\section{C. Branching process model}

We model injection avalanches as 
a compound of a random number $M$ of independent 
sub-avalanches, each described by a Galton-Watson (GW) branching process 
\cite{Harris}. The total avalanche size is then
\begin{equation} \label{sum_of_independent_trees}
S = \sum_{m=1}^M Y_m
\end{equation}
in which $Y_m$ is the size (namely, the number of branches of the GW process) of the $m$-th sub-avalanche ($Y_m$ is called $S_m$ in the main text. In 
this Section, we change notation to avoid confusion with $S$.)
$M$ is the random number of pairs that were destabilized by the initial injection and that
later relaxed, with $\langle M \rangle=\rho_L$.

 Each hop is assumed to create $X$ subsequent hops with probability $p(X)$,
with $\langle X\rangle=\lambda_L$, where we allow the mean to depend 
on $L$ to account for finite-size effects.
The sub-avalanche size is thus $Y_m=\sum_{k=1}^\infty Z_k$, 
where $Z_k$ is the number of hops at generation $k$ 
of the branching tree, which satisfies the recursion relation
\beq
Z_1=1, \qquad Z_{n+1}=\sum_{i=1}^{Z_n} X_i
\eeq
where the $X_i$ are independent, identically distributed (i.i.d.) copies of $X$. 

The probability distributions of $Y$, denoting with this symbol any 
of the i.i.d. variables $Y_m$, and of $S$ 
can be determined with generating function methods.
It is easy to show (see e.g. Ref.\onlinecite{Harris}) 
that the generating functions of $Y$ and $S$,
$g_Y(t)=\langle t^Y\rangle$ and $g_S(t) = \langle t^S\rangle$, satisfy
the equations 
\beq
g_Y(t) = t g_X(g_Y(t))
\label{gY}
\eeq
\beq
g_S(t) = g_M(g_Y(t))
\label{gS}
\eeq 
where $g_X(t) = \langle t^X \rangle$ and $g_M(u)=\langle u^M\rangle$
are the generating functions of $X$ and $M$, respectively.

By expanding $g_Y(t)$ in $t$ one immediately obtains $\langle Y\rangle=(1-\lambda_L)^{-1}$.
From Eq.(\ref{gY}) one can also obtain \cite{Harris} the 
asymptotic behavior of $p(Y)$ at large $Y$,
\beq
p(Y)\sim Y^{-3/2} e^{-Y/S_c},
\eeq
where the cutoff diverges as  $S_c \sim \langle X^2\rangle / (1-\lambda_L)^2$ for $\lambda_L \to 1^-$,
as follows from the general expression given by Harris \cite{Harris} for $S_c$.
The same result was also obtained by Fisher with a similar method \cite{Fisher}.

In the special case in which  $X$ is Poisson distributed, we have $g_X(t) = \exp(\lambda_L (x-1))$
and Eq.(\ref{gY}) can be solved explicitly, giving 
\beq
g_Y(t) = - \frac{1}{\lambda_L} W \left[ - \lambda_L \exp(-\lambda_L) t \right] \, ,
\eeq
where $W(x)$ is the Lambert $W$ function, with series
representation
$
W(x) = \sum_{n=1}^\infty (-n)^{n-1} x^n/n! $ for $|x|<\exp(1)$.
This immediately gives 
\beq
p(Y) = \frac{{(\lambda_L Y)}^{Y-1}}{Y!} \exp[- \lambda_L Y]  \, .
\label{pY}
\eeq
The asymptotic behavior of Eq.(\ref{pY}) for large $Y$ is
\beq
p(Y)\sim 
\frac{1}{\sqrt{2 \pi \lambda_L^2}} Y^{-3/2} \exp[- Y/S_c] \label{probability_Y_sim} 
\eeq
with 
\beq
S_c = (\lambda_L - 1 - \log \lambda_L )^{-1}.
\label{YcPoisson}
\eeq 

If $M$ is also Poisson-distributed with mean $\rho_L$, 
by expanding Eq.(\ref{gS}) around $t=0$ we obtain 
\beq
p(S) =
\frac{\rho_L (\lambda_L S + \rho_L)^{S-1} }{S!} \exp[-(\lambda_L S + \rho_L)]  \,.
\label{pSexact}
\eeq
Eq.(\ref{pSexact}) fits very well the data for injection avalanches, as shown 
in Fig.3 of the main text. The fit parameters $\rho_L$ and $\lambda_L$ for
each $L$ are consistent with linear scaling $\rho_L =0.09(1) L$ and $S_c = 0.31(4) L$,
for $S_c$ given by Eq.(\ref{YcPoisson}).

For $S\ll \rho_L$,  $p(S)$ in Eq.(\ref{pSexact}) 
tends to a Poisson distribution, while 
for $S\gg \rho_L$ it has the form of a power law with 
exponential cutoff 
\beq
p(S) \sim B_L \, S^{-3/2} \exp{ \left[ - S/S_c  \right] } \label{pSa} \, .
\eeq
where $B_L=\rho_L \exp(\rho_L / \lambda_L -\rho_L) (2 \pi \lambda_L^2)^{-1/2}$. 
A scaling plot according to this form is shown in Fig.3 of the main text.

The function $p(S)$ in Eq.(\ref{pSexact}) presents a maximum at $S=S^*$. To estimate $S^*$,
we solve $d\log p(S)/dS =0$ using Stirling's approximation, which gives
\beq
\log\frac{\lambda_L S^*+\rho_L}{S^*}+\frac{\lambda_L (S^*-1)}{\lambda_L S^*+\rho_L}-\lambda_L
-\frac{1}{2S^*}=0 \, .
\eeq
To leading order we then have $S^*=\rho_L /(1-\lambda_L)=\langle S \rangle$ and
 expanding in the small parameter $x=[\rho_L (1-\lambda_L)]^{-1}\sim \sqrt{S_c}/\rho_L \sim 1/\sqrt{L}$ 
we obtain the corrections 
\beq
S^*=\langle S\rangle (1-\lambda_L x - \frac{\lambda_L}{2} x^2 + O(x^3))\, .
\eeq
The important point to note is that in 3D, since $\rho_L\propto W^{1/\delta-1} L$ and 
$S_c\propto L W^{-1/\delta}$, as discussed in the main text, and 
since $\langle S\rangle \sim \rho_L S_c^{1/2}$, the ratio $S^* / S_c$ is proportional
to $L^{1/2}/W^{\frac{3}{2 \delta}-1}$ and thus diverges for large $L$.
 Hence, the power law behavior $p(S)\sim S^{-3/2}$ is never observed, since the 
position of the maximum is increasingly larger than the cutoff. A similar estimate
in 2D, noting that $\rho_L$ and $S_c$ both increase logarithmically with $L$, 
gives $S^* / S_c \sim (\log L)^{1/2}$.

\section{D. Tests of the branching process model}

\subsection{D.1. Procedure}

In order to test the branching process description of the avalanches,
we reconstructed the genealogical tree of each individual injection avalanche
by keeping track of the offspring of each hop, as described in the following.

Let $J_1, J_2, \dots, J_S$
be the hops taking place in the course of an avalanche
of size $S$, in the order they occurred. We call
$\uss_1$  the set of electron-hole pairs
that become unstable after the initial electron injection, 
and $\uss_i$, $i>1$  the set of pairs that are unstable 
after hop $J_{i-1}$  and before hop $J_i$ (regardless of when they
became unstable). 

We call {\it progeny} of a hop $J_i$ the set of hops $J_l$ with $l>i$
such that
$J_l \in \uss_k $ for all $k=i+1,\dots,l$ and $J_l \notin \uss_i$,
namely those hops that relaxed pairs that were destabilized 
by $J_i$
and remained unstable at all times until  they relaxed. (The progeny
does not include pairs that, after being destabilized by $J_i$, are stabilized again by subsequent hops without relaxing.) We call $X_i$ the size of the progeny of $J_i$
and measure its distribution 
$p(X)=\langle S^{-1}  \sum_{i=1}^S \delta_{X_i,X}\rangle$
by averaging over all the avalanches (one avalanche per disorder realization
at $K=1/2$). 

The {\it first generation} of an avalanche is the set of hops $J_i$ such
that
$J_i \in \uss_k $ for all  $k=1,\dots,i$, namely those hops
that relaxed pairs that were destabilized by the initial injection
and remained unstable at all times until they relaxed. We call $M$ the number
of first-generation hops, and denote by $p(M)$ its distribution, again estimated
by averaging over all avalanches. 

The {\it $k$-th generation} of the avalanche is the set of hops 
that belong to the progeny of a hop that itself belongs to generation $k-1$.  
Finally, we call {\it sub-avalanche} each of the $M$ first generation hops 
together with its progeny (i.e. the subtree arising from it).

\subsection{D.2. Statistics of the first generation}

The top two panels of Fig.~\ref{fig:par_est} summarize the statistics of $M$ obtained using
the above procedure for $W=2$.
Panel (a) shows that the cumulative distribution function of $M$ 
for different system sizes agrees reasonably well with a Poisson distribution 
of mean $\rho_L$ equal to the measured mean $\langle M\rangle$. The deviation from
the Poisson distribution 
reflects the correlations
between soft electron-hole pairs, and can be quantified by the relative difference
 between $\rho_L$ and 
the measured variance of $M$, both shown in the panel (b), 
which is around 30$\%$. 

The mean grows approximately linearly with $L$, the best fit giving $\rho_L = 0.068(6) L$,
and is not far from the estimate $\rho_L$ obtained from the two-parameter fits of 
Eq.(\ref{pSexact}) to $p(S)$ for injection avalanches,
also shown in Fig.~\ref{fig:par_est} (b),
which is fitted by $\rho_L = 0.09(1) L$ as discussed in Section C.
The linear increase of $\rho_L$ with $L$ agrees with the mean-field estimate in Eq.(3) of the
main text, which assumes a constant pair DOS. In fact, it is known that
at small $\omega$ the pair DOS decreases logarithmically \cite{baranovski80},
which produces logarithmic corrections to the linear dependence of $\rho_L$ on $L$. 
Including these corrections, we reproduce the deviation from linearity observed in Fig.~\ref{fig:par_est}.

\subsection{D.3. Statistics of the branching ratio}

\begin{figure}[h!]
\includegraphics[height=0.98\linewidth,width=1.19\linewidth,angle=-90]{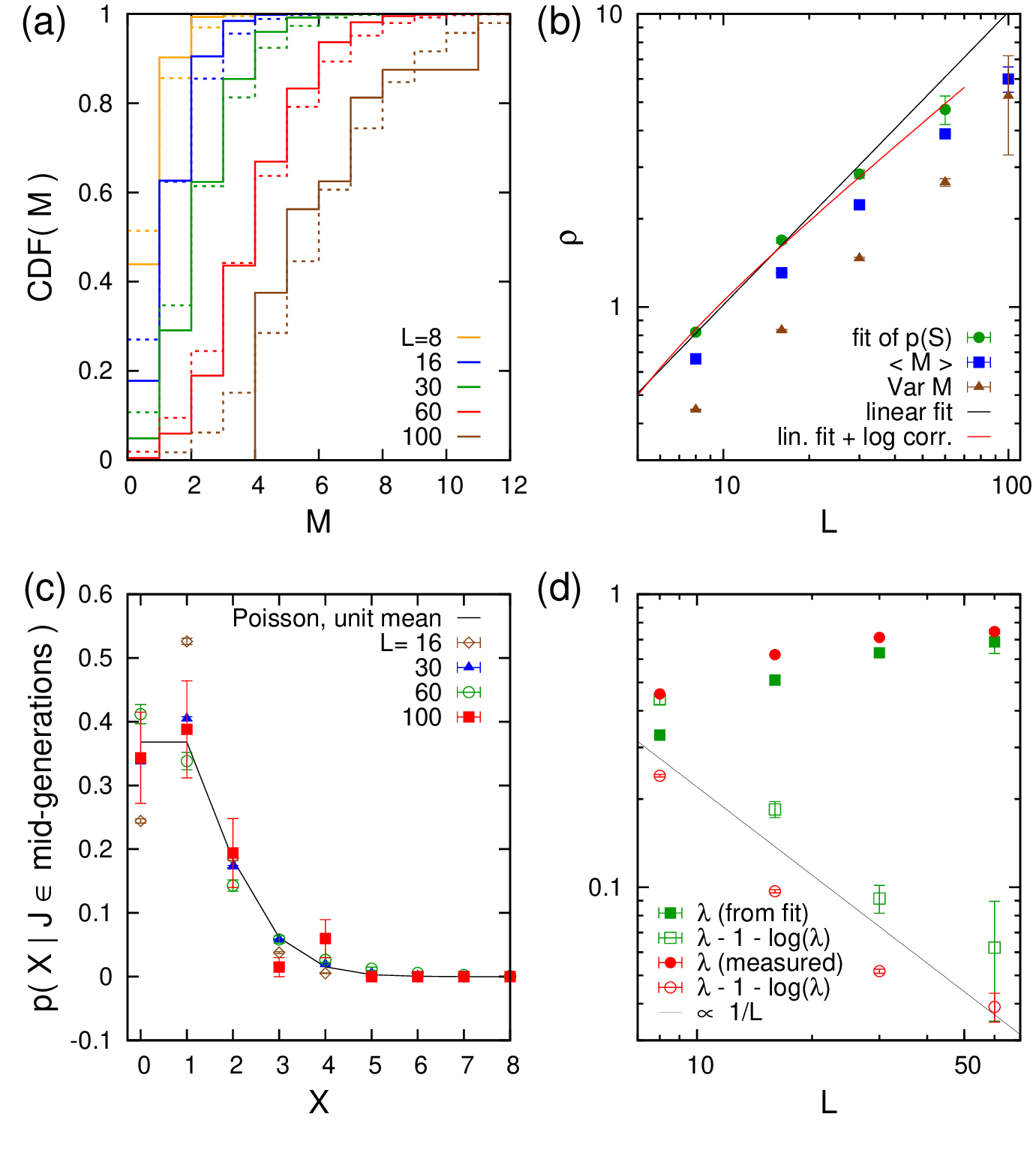}
\caption{Statistics of the reconstructed avalanche tree for $W=2$.
{\bf (a)} The cumulative distribution function of the number $M$ of first-generation hops  (continuous lines) 
 is compared with a Poisson distribution of mean $\rho_L=\langle M\rangle$ (dashed lines) for 
 different sizes $L$.
{\bf (b)} $\la M \ra$ (blue squares) compared to the 
value of $\rho_L$ (green circles) obtained by fitting Eq.(\ref{pSexact}) to the
size distribution of injection avalanches in 3D, shown in Fig.3 of the main text.
The two values are fairly close, supporting the branching
process description, and increase linearly with $L$, possibly with some
logarithmic corrections. Also shown is the variance of $M$ (brown triangles), 
which differs from the mean, highlighting a deviation from Poisson statistics. 
{\bf (c)} The  distribution of the number of offspring $X$, measured in 
mid-generations
is approximately independent of $L$ and well described by a Poisson 
distribution with mean $\lambda_L=1$. 
{\bf (d)} The value of $\lambda_L$ obtained by fitting Eq.(\ref{pSexact}) to the
data of Fig.3 of the main text  is compared with $\lambda_L=\la X \ra$ 
measured including all generations.} 
\label{fig:par_est}
\end{figure}

The statistics of $X$ for $W=2$ are summarized in the bottom panels of Fig.~\ref{fig:par_est}.
The distribution $p(X)$ obtained by averaging over all hops (not shown)
shows a dependence on the system size. In particular, as shown in 
the panel (d),
its mean $\lambda_L = \langle X \rangle$
increases with $L$, giving a linear increase $S_c=0.45(9) L$ if we define
$S_c$ as in Eq.(\ref{YcPoisson}). 
This agrees very well with the observed linear increase of the cutoff of $p(S)$
for displacement avalanches, which can be fitted to $S_c = 0.46(8) L$ (Fig.1 of the main text), 
and fairly well with the analogous fit for injection avalanches which gives 
$S_c = 0.31(4) L$ (Fig.3 of the main text).

We ascribe the dependence of $p(X)$ on the system size to the fact that
when an avalanche reaches the boundary of the system and enters again from the ``opposite 
side'', it revisits regions already affected by the avalanche, which are more stable and thus
tend to reduce $\la X \ra$. 
To filter out this finite-size effect and test the validity of the branching process 
description, we also estimated $p(X)$ including only ``mid-generations'' hops, namely
hops that {\it a)} belong to sub-avalanches that are at least six generations deep and {\it b)}
are within the central $40\%$ of the generations of the sub-avalanche to which
they belong. As shown in Fig.~\ref{fig:par_est} (c) the distribution measured in this way is 
relatively independent of $L$. For the larger system sizes $L=60,100$, for which 
the mid-generation protocol should filter out finite-size effects more effectively,
$p(X)$ is well described by a Poisson distribution of mean unity, in agreement
with the assumption of a critical GW process.

\subsection{D.4. Size distribution of sub-avalanches}

\begin{figure}[h!]
\includegraphics[height=0.9\linewidth,width=0.65\linewidth,angle=-90]{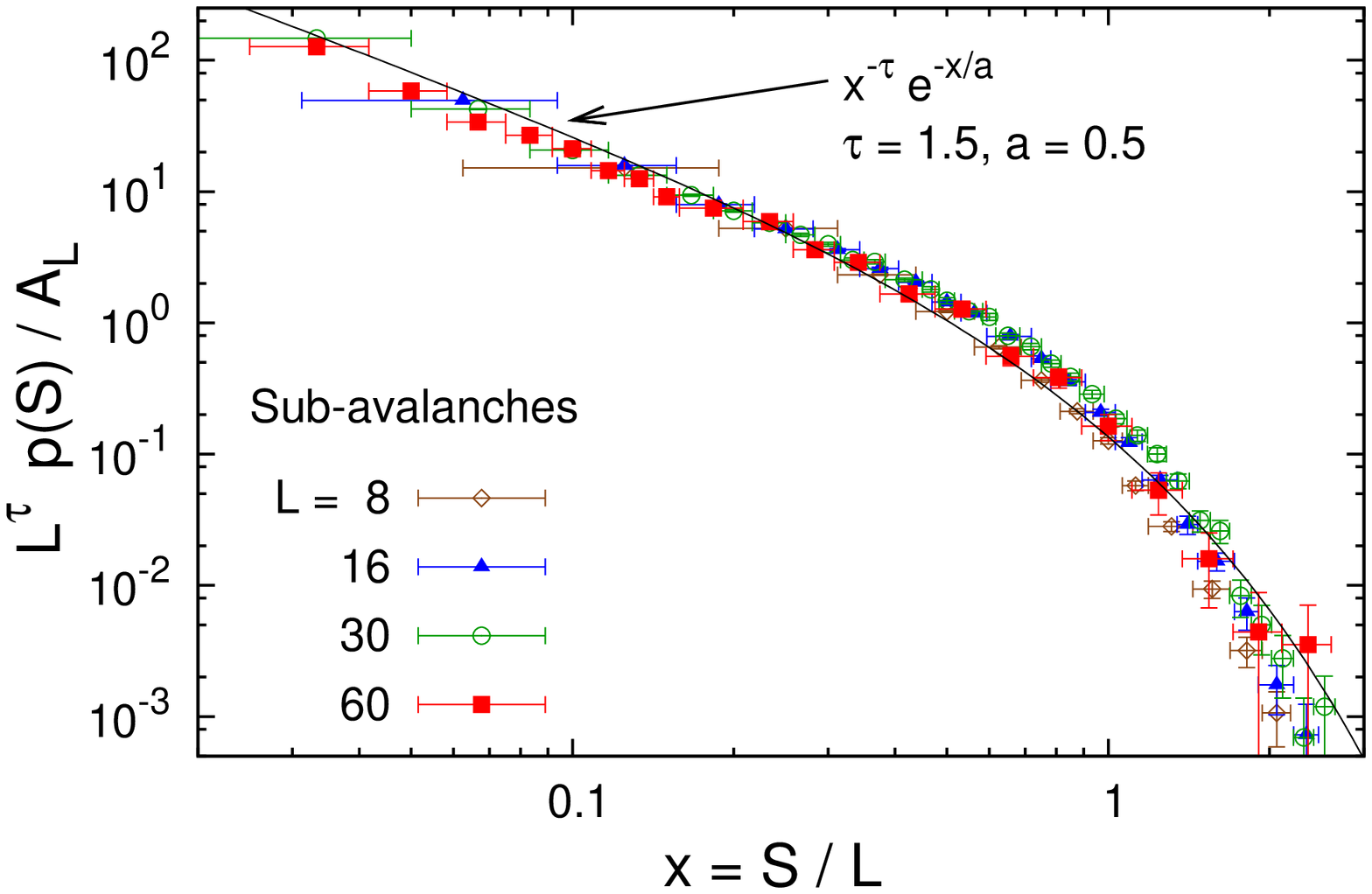}
\caption{Size distribution of the sub-avalanches of
injection avalanches, obtained by identifying the
progeny of the hops that were destabilized by the electron injection. }
\label{fig:subav}
\end{figure}

Fig.\ref{fig:subav} shows the size distribution of the sub-avalanches,
defined as explained in paragraph D.1.
The shape of $p(S)$ is practically
indistinguishable from that of displacement avalanches shown in 
Fig.1 of the main text, supporting our assumption that injection avalanches
are well approximated by the compound of $M$ displacement avalanches.

\section{E. Avalanches in two dimensions}

\begin{figure}[h]
\includegraphics[height=\linewidth,width=0.9\linewidth,angle=-90]{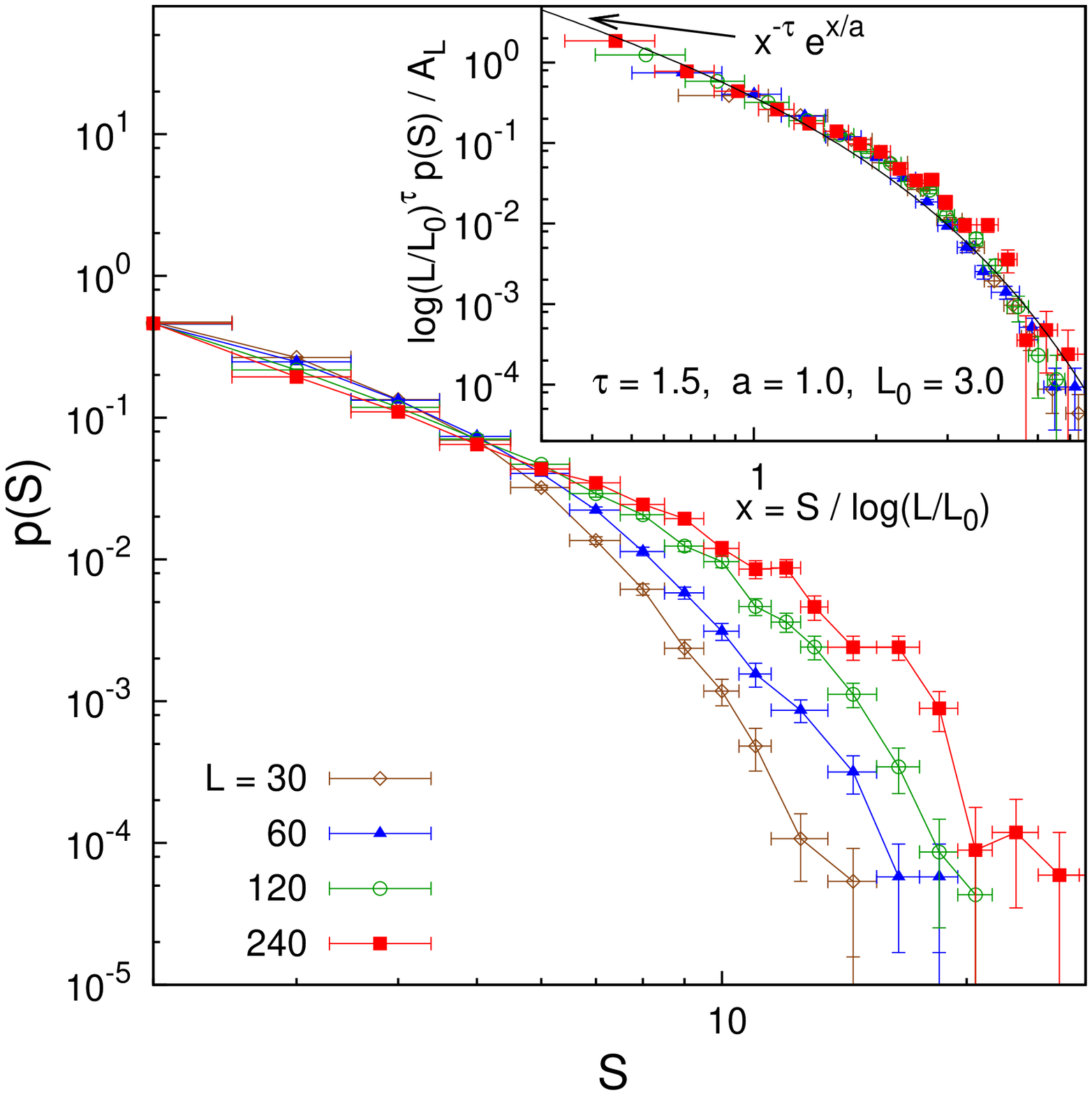}
\caption{Avalanche size distribution for displacement avalanches in 2D,
evolved with the $\xi=\infty$ dynamics, for  $W = 2$, $K=1/2$, and
$L=30, 60, 120, 240$. 
Inset: rescaling of the data according to Eq.(2) of the main text,
with a logarithmic cutoff. }
\label{fig:DTA_2D}
\end{figure}

We report here the results of our simulations in two dimensions,
which followed the same protocol described in the main text for three dimensional
simulations.
Fig.\ref{fig:DTA_2D} shows the size distribution of displacement avalanches
under the $\xi=\infty$ dynamics. The data can be fitted by Eq.(2) of the main text
with $\tau=1.5$, similarly to the results in 3D (Fig.1 of the main text). However in 
2D the cutoff increases logarithmically with $L$ instead of linearly, as shown
by the scaling plot in the figure inset.  

\begin{figure}[h!]
\includegraphics[height=\linewidth,width=0.9\linewidth,angle=-90]{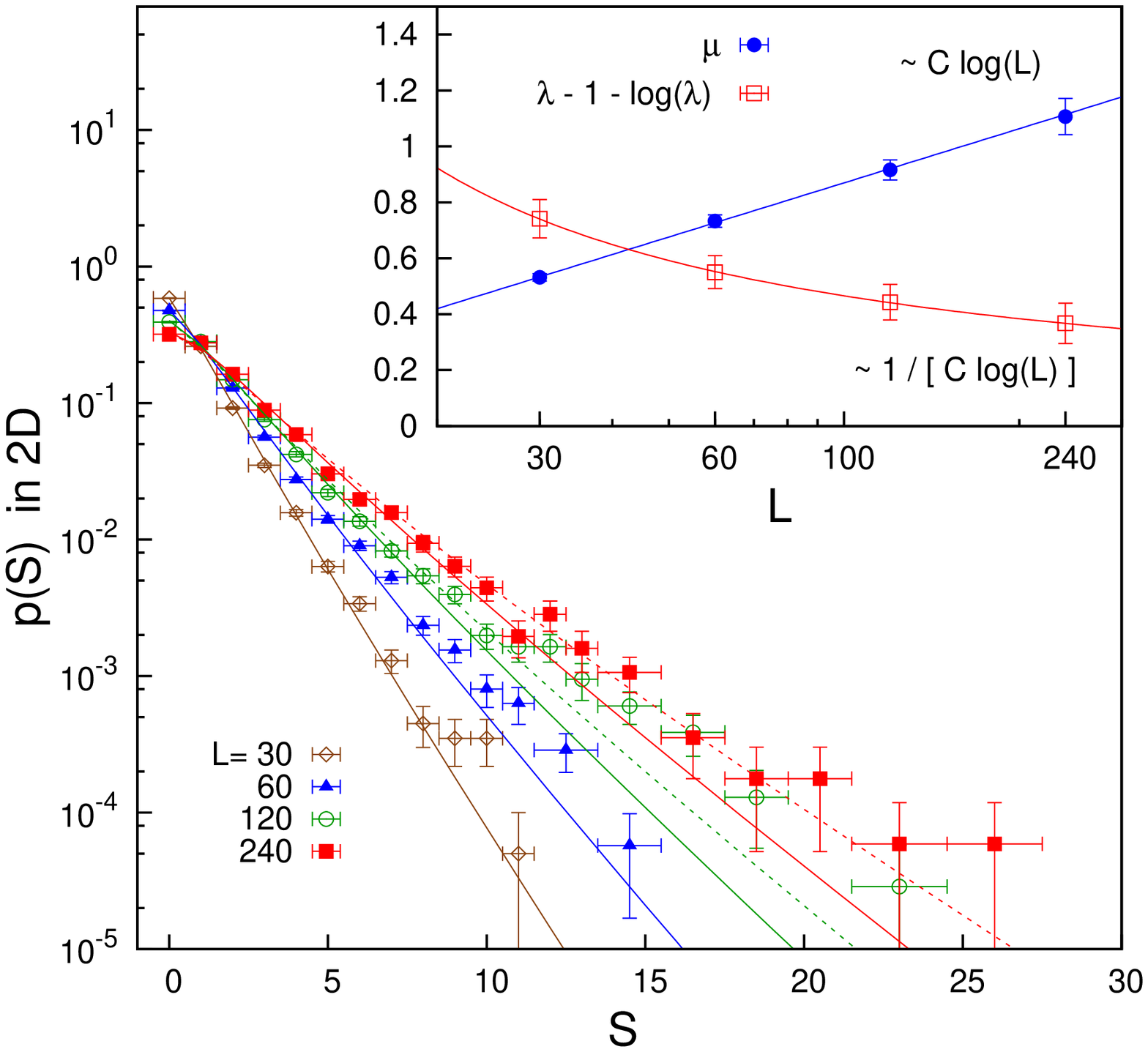}
\caption{Same as Fig.~\ref{fig:DTA_2D} except that avalanches are triggered 
by injection.  
The solid lines are fits to Eq.\ref{pSexact}. The inset shows that the 
$L$-dependence of both fit parameters $\lambda_L$ and $\rho_L$ is consistent
with a logarithmic dependence $S_c,\rho_L \propto \log L$.}
\label{fig:CTA_2D}
\end{figure}

Fig.\ref{fig:CTA_2D} shows the size distribution of injection avalanches in 2D under the
same dynamics.
As in displacement avalanches, the cutoff grows with $L$ much more slowly than in 3D
(see Fig.3 of the main text), and is consistent with a logarithmic dependence. 
$p(S=0)$ decreases much more slowly with $L$ than in 3D,
in agreement with the estimate $p(S=0)=e^{-\langle M\rangle}$ and with a logarithmic growth of $\langle M\rangle$ with
$L$ implied by the mean-field estimate in Eq.(3) of the main text.
By fitting Eq.(\ref{pSexact}) to the data in Fig.~\ref{fig:CTA_2D}, we determine
the parameters $\lambda_L$ and $\rho_L$. As shown in the figure inset, 
the fitted parameters are consistent with a logarithmic growth $\rho_L \propto \log L$
and $S_c(\lambda_L) \propto \log L$.

\vspace{1cm}
\bigskip

\end{appendices}


\begin{thebibliography}{99}
\bibitem{foot}
The number of sites destabilized by an electron exchange
is of order
$\int_a^L dr r^{d-1} \int_0^{e^2/r} g(\e) d\e \sim L^{d-\delta-1}$ if
$g(\e)\sim |\e|^\delta$. In order to reach a stable configuration,
after the last exchange this number cannot diverge, thus $\delta\geq d-1$.
\bibitem{example} The percolation approach gives 
  $r_{typ}\simeq (\xi/4)(T_0/T)^{1/2}$ with $T_0 \simeq 2.8 \,e^2/\xi$,
 with an associated time scale $t\simeq \tau_0 e^{-2 r_{typ}/\xi}$ where $\tau_0 \simeq 10^{-12}$s
\cite{ESbook}.
For $\xi\simeq a$ and $T \simeq 10^{-3} e^2/a$
we have $r_{typ} \simeq 13 \, a$. Extrapolating from
Fig.S1 \cite{Supp} and Fig.3, single
electrons will trigger avalanches with $O(10^2)$ hops with 
probability $O(10^{-3})$  and duration $t \simeq 0.2\,$ s.
\end{thebibliography}
\end{document}